\documentclass[12pt,reqno]{article}
\usepackage{amsfonts}
\usepackage{amsmath}
\usepackage{amsbsy}

\usepackage{amssymb,latexsym}
\numberwithin{equation}{section}

\begin{document}

\title{Plane Symmetric Solutions of Gravitational Field Equations in Five Dimensions
}
\begin{titlepage}
\author{ A. N. Aliev\footnote{E.mail: aliev@gursey.gov.tr} \\
{\small Feza G\"{u}rsey Institute, 34684 \c{C}engelk\"{o}y,
\.{I}stanbul, Turkey} \\ \\
H. Cebeci\footnote{E.mail:
cebeci@gursey.gov.tr} \\
{\small Department of Physics, Anadolu University, 26470 Eski\c{s}ehir, Turkey} \\ \\
T. Dereli\footnote{E.mail: tdereli@ku.edu.tr}\\{\small Department
of Physics,  Ko\c{c} University, 34450 Sar{\i}yer-\.{I}stanbul,
Turkey} }

\date{ }

\maketitle

\bigskip

\begin{abstract}
\noindent We present the effective field equations obtained from a
generalized gravity action with Euler-Poincar\'{e} term and a
cosmological constant in a $D$ dimensional bulk space-time. A
class of plane-symmetric solutions  that describe a 3-brane world
embedded in a $D=5$ dimensional bulk space-time are given.
\end{abstract}

\bigskip

\noindent {\small Paper presented at the {\bf 5th Workshop on
Quantization, Dualities and Integrable Systems}, Pamukkale
University, 23-28 January 2006, Denizli, Turkey.}

\end{titlepage}

\section{Introduction}

Brane-world theories that  receive a lot of interest recently  are
strictly motivated by string models \cite{horava-witten}. They
were mainly proposed to provide new solutions to the hierarchy
problem and compactification of extra dimensions
\cite{Arkani},\cite{randall}. The main content of the brane-world
idea is that we live in a four dimensional world embedded in a
higher dimensional bulk space-time. According to the brane-world
scenarios, the gauge fields, fermions and scalar fields of the
Standard Model should be localised on a 3-brane, while gravity may
freely propagate into the higher dimensional bulk.

In our previous work \cite{dereli5} we  derived covariant
gravitational field equations on a 3-brane embedded in a
five-dimensional bulk space-time with $\mathbb{Z}_{2}$ symmetry in
a generalization that included a dilaton scalar as well as the
second order Euler-Poincar\'{e} density in the action.  We
introduced a general ADM-type coordinate setting to show that the
effective gravitational field equations on the 3-brane remain
unchanged, however, the evolution equations off the brane are
significantly modified due to the acceleration of normals to the
brane surface in the non-geodesic, ADM slicing of space-time.

In the second part of this paper, using the language of
differential forms, we present the field equations of a
generalized gravity model with a dilaton 0-form and an axion
3-form in Einstein frame from an action that includes the second
order Euler-Poincar\'{e} term and a cosmological constant  in a
D-dimensional bulk space-time.  In the third part, we present some
plane-symmetric solutions  that generalize the well-known
domain-wall solution \cite{lukas}.

\section{Model}

We consider a $D$-dimensional bulk space-time manifold $M$
equipped with a metric $g$ and a torsion-free, metric compatible
connection $\nabla$. We determine our gravitational field
equations by a variational principle from a $D$-dimensional action
that includes the second order Euler-Poincar\`{e} term and a
cosmological constant
\begin{equation}
I[e, \omega, \phi, H] = \int_M {\cal L}
\end{equation}
where in the Einstein frame the Lagrangian density D-form
\cite{dereli4}
\begin{eqnarray}
{\cal L} &=& \frac{1}{2} R^{ab} \wedge \ast (e_{a} \wedge e_{b} ) - \frac{\alpha}{2} d \phi \wedge \ast d \phi +
\frac{\beta}{2} e^{ - \beta_{2} \phi} H \wedge \ast H + \Lambda e^{ - \beta_{1} \phi} \ast 1  \nonumber \\
& & + \frac{\eta}{4} R^{ab} \wedge R^{cd} \wedge \ast ( e_{a}
\wedge e_{b} \wedge e_{c} \wedge e_{d} )   \nonumber \\
 & & + ( d e^{a} + \omega^{a}\,_{b} \wedge e^{b} ) \wedge \lambda_{a} + ( d H -
\frac{ \varepsilon }{2} R_{ab} \wedge R^{ab} ) \wedge \mu \; .
\label{22}
\end{eqnarray}
Here $\lambda_a$ and $\mu$ are Lagrange multiplier forms that upon
variation impose the zero-torsion  and anomaly-freedom
constraints.

The final form of the variational  field equations to be solved
are the Einstein field equations
\begin{eqnarray}
\frac{1}{2} R^{ab} \wedge \ast ( e_{a} \wedge e_{b} \wedge e_{c} ) = - \frac{ \alpha }{2}  \tau_{c} [\phi] +
\frac{\beta}{2} e^{ - \beta_{2} \phi} \tau_{c} [H] - \Lambda e^{ - \beta_{1} \phi} \ast e_{c} \nonumber \\
- \frac{\eta}{4} R^{ab} \wedge R^{dg} \wedge \ast ( e_{a} \wedge e_{b} \wedge e_{d} \wedge e_{g} \wedge e_{c} ) \nonumber \\
- 2 \varepsilon \beta D ( e^{ - \beta_{2} \phi} \iota_{b} ( R^{b}\,_{c} \wedge \ast H ) ) -
\frac{\varepsilon \beta}{2} e_{c} \wedge D ( e^{ - \beta_{2} \phi} \iota_{s} \iota_{l} ( R^{ls} \wedge \ast H ) ) ,
\label{33}
\end{eqnarray}
where the dilaton stress-energy forms
\begin {equation}
\tau_{a} [\phi] = \iota_ad\phi *d\phi + d\phi \wedge \iota_a
*d\phi \nonumber
\end{equation}
and the axion stress-energy forms
\begin{equation}
\tau_{a} [H] = \iota_a H \wedge *H + H \wedge \iota_a *H \; ,
\nonumber
\end{equation}
the dilaton scalar field equation
\begin{equation}
\alpha d ( \ast d \phi ) = \frac{\beta_{2} \beta}{2}  e^{ - \beta_{2} \phi} H \wedge \ast H + \Lambda \beta_{1}  e^{ - \beta_{1} \phi} \ast 1 ,
\label{34}
\end{equation}
and the axion field equations
\begin{equation}
d H = \frac{\varepsilon}{2} R_{ab} \wedge R^{ab} \quad , \quad d
(e^{- \beta_{2} \phi } \ast H ) = 0  . \label{36}
\end{equation}

\section{Plane symmetric solutions in $D=5$}

We investigate below a class of plane symmetric solutions  in
5-dimensions. We consider the metric
\begin{equation}
g = - f^{2} ( t,\omega ) d t^{2} + u^{2} ( t,\omega ) d \omega^{2}
+ g^{2} ( t,\omega ) \left( \frac{d x^{2} + d y^{2}
+ d z^{2}}{\left( 1 + \frac{k r^{2} }{4} \right)^{2}} \right) ,
\label{37}
\end{equation}
the dilaton scalar field
\begin{equation}
\phi = \phi ( t,\omega )
\label{38}
\end{equation}
and 3-form gauge field
\begin{equation}
H = h ( t,\omega ) \frac{d x \wedge d y \wedge d z}{\left( 1 +
\frac{k r^{2} }{4} \right)^{3}} \label{39}
\end{equation}
in terms of local coordinates
$$
x^{M} : \left\{ x^{0} = t, x^{5} = \omega , x^{1} = x, x^{2} = y,
x^{3} = z \right\}  .
$$
We choose our co-frame 1-forms as
\begin{equation}
e^{0} = f ( t,\omega ) d t, \qquad e^{5} = u ( t,\omega ) d \omega , \qquad e^{i} =
g ( t,\omega ) \frac{d x ^{i}}{\left( 1+ \frac{k r^{2}}{4} \right)}, \quad i = 1,2,3  .
\label{40}
\end{equation}
Then we calculate the Levi-Civita connection 1-forms
\begin{equation}
\omega^{0}\,_{i} = \frac{g_{t}}{f g} e^{i} , \qquad \omega^{i}\,_{j} = \frac{k}{2 g} ( x^{i} e^{j} - x^{j} e^{i} ) ,
\label{41}
\end{equation}
\begin{equation}
\omega^{0}\,_{5} = \frac{u_{t}}{f u} e^{5} + \frac{f_{\omega}}{f u} e^{0}, \qquad \omega^{i}\,_{5} = \frac{g_{\omega}}{u g} e^{i} .
\label{42}
\end{equation}
and the corresponding curvature 2-forms
\begin{equation}
R^{ij} = \frac{1}{g^{2}} \left\{ k + \left( \frac{g_{t}}{f} \right)^{2} - \left( \frac{g_{\omega}}{u} \right)^{2} \right\} e^{i} \wedge e^{j} ,
\label{43}
\end{equation}
\begin{equation}
R^{05} = \frac{1}{f u} \left\{ \left( \frac{f_{\omega}}{u} \right)_{\omega} - \left( \frac{u_{t}}{f} \right)_{t} \right\} e^{5} \wedge e^{0} ,
\label{44}
\end{equation}
\begin{equation}
R^{0i} = \frac{1}{f g} \left\{ \left( \frac{g_{t}}{f} \right)_{t} - \frac{f_{\omega} g_{\omega} }{u^{2}} \right\} e^{0} \wedge e^{i}
+ \frac{1}{ug} \left\{ \left( \frac{g_{t}}{f} \right)_{\omega} - \frac{u_{t} g_{\omega} }{f u} \right\} e^{5} \wedge e^{i} ,
\label{45}
\end{equation}
\begin{equation}
R^{i5} = \frac{1}{f g} \left\{ \frac{f_{\omega} g_{t} }{f u} - \left( \frac{g_{\omega}}{u} \right)_{t} \right\} e^{i} \wedge e^{0}
+ \frac{1}{ug} \left\{ \left( \frac{g_{\omega}}{u} \right)_{\omega} - \frac{g_{t} u_{t}}{f^{2}} \right\} e^{5} \wedge e^{i} .
\label{46}
\end{equation}
From these expressions we note that $ R_{ab} \wedge R^{ab} = 0 $.
Therefore $d H = 0$ implying that
\begin{equation}
H = \frac{Q}{g^{3}} e^{1} \wedge e^{2} \wedge e^{3}
\label{47}
\end{equation}
where $Q$ may be identified as a magnetic charge. Now, for
simplicity, we let $k=0$ and take the functions $g$, $f$ and $u$
independent of time. Then we obtain the following system of
coupled ordinary differential  equations (${ }^{\prime}$ denotes
derivative with respect to $\omega$):
\begin{eqnarray}
2 G - 2 C - B - A &=& - \eta ( 2 C G - A B ) - \frac{\alpha}{2} \left( \frac{\phi^{\prime}}{u} \right)^{2} \nonumber \\
& & - \frac{\beta}{2} \frac{Q^{2}}{g^{6}} e^{ - \beta_{1} \phi} + \Lambda e^{ - \beta_{2} \phi } ,
\label{48}
\end{eqnarray}
\begin{equation}
3 A - 3 G = 3 \eta G A + \frac{ \alpha }{2} \left( \frac{ \phi^{\prime} }{u} \right)^{2}
- \frac{\beta}{2} \frac{Q^{2}}{g^{6}} e^{ - \beta_{2} \phi} - \Lambda e^{ - \beta_{1} \phi} ,
\label{49}
\end{equation}
\begin{equation}
3 C + 3 A = - 3 \eta C A - \frac{ \alpha }{2} \left( \frac{\phi^{\prime}}{u} \right)^{2}
 - \frac{\beta}{2} \frac{Q^{2}}{g^{6}} e^{ - \beta_{2} \phi} - \Lambda e^{ - \beta_{1} \phi} ,
\label{50}
\end{equation}
\begin{equation}
\alpha \left( \frac{ \phi^{\prime} f g^{3} }{u} \right)^{\prime} \frac{ 1 }{g^{3} f u} =
\frac{\beta_{2} \beta }{2} e^{ - \beta_{2} \phi} \frac{Q^{2}}{g^{6}} + \Lambda \beta_{1} e^{ - \beta_{1} \phi} .
\label{51}
\end{equation}
where
\begin{equation}
 A = - \left( \frac{ g^{\prime} }{g} \right)^{2} \frac{ 1 }{u^{2}} \qquad \qquad B = - \left( \frac{f^{\prime}}{u} \right)^{\prime} \frac{ 1 }{f u} ,
\label{52}
\end{equation}
\begin{equation}
C = - \frac{ f^{\prime} g^{\prime} }{u^{2} f g} \qquad \qquad G = \left( \frac{ g^{\prime} }{u} \right)^{\prime} \frac{ 1 }{u g} .
\label{53}
\end{equation}

We will give below some  special classes of solutions:
\bigskip

\noindent {\bf Case:} $\phi=constant$, $H=0$ and $\eta = 0$.

\bigskip

\noindent Here the Euler-Poincar\'{e} term is absent, $H=0$ and
the dilaton scalar is constant. We obtain the AdS solution in
5-dimensions that is also known as Randall-Sundrum model
\cite{randall}:
\begin{equation}
g = d \omega^{2} + e^{\mp 2 p \omega } ( - d t^{2} + d x^{2} + d y^{2} + d z^{2} ) .
\label{54}
\end{equation}
where $p^{2} = \frac{ \Lambda }{6}$.
\bigskip

\noindent {\bf Case:} $\phi=constant$, $H=0$.

\bigskip

\noindent Here $H=0$ and the dilaton scalar is constant. Solutions
are  given by the metric
\begin{equation}
g = d \omega^{2} + e^{ \mp 2 s \omega } ( - d t^{2} + d x^{2} + d y^{2} + d z^{2} )
\label{55}
\end{equation}
where
\begin{equation}
s^{2} =  \frac{ 1 + \sqrt{ 1 - \frac{ \eta \Lambda }{3} } }{\eta}
\label{56}
\end{equation}
provided that $ \Lambda \eta \leq 3 $.
When $ \eta \Lambda = 3 $, the solution may alternatively be given in AdS form as
\begin{equation}
g = - 4 \cosh^{2} ( l \omega )d t^{2} + d \omega^{2} + 4 \sinh^{2} ( l \omega ) ( d x^{2} + d y^{2} + d z^{2} )
\label{57}
\end{equation}
where $l^{2} = \frac{1}{\eta} $.
\bigskip

\noindent {\bf Case:} $\eta = 0$, $H=0$.

\bigskip

\noindent Here the Euler-Poincar\'{e} term is absent and $H=0$. We
obtain the following solution:
\begin{equation}
g = e^{\frac{16 \alpha}{3 \beta_{1} } \phi(\omega)} d \omega^{2} + e^{\frac{4 \alpha}{3 \beta_{1} }\phi(\omega)} ( - dt^{2}
+ d x^{2} + d y^{2} + d z^{2} )
\label{58}
\end{equation}
with
\begin{equation}
\phi(\omega) = \frac{1}{\left( \frac{\beta_{1}}{2} - \frac{8 \alpha}{3 \beta_{1}} \right)} \ln \left\vert \sqrt{ \frac{2 \beta_{1} \Lambda }{\left( \frac{ 16 \alpha }{3 \beta_{1} } - \beta_{1} \right) \alpha } }\left( \frac{ \beta_{1} }{2} - \frac{8 \alpha }{3 \beta_{1} } \right) \omega + C_{0} \right\vert
\label{59}
\end{equation}
where $C_{0}$ is an integration constant. When $\beta_{1} = 2 $, it reduces to a supersymmetric domain wall solution presented in \cite{lukas}.
\bigskip

\noindent {\bf Case:} $\eta = 0$.

\bigskip

\noindent In this case the solution possesses a magnetic charge.
It is given by
\begin{equation}
g = e^{ \frac{4 ( \beta_{1} - \beta_{2} )}{3} \phi(\omega) } d \omega^{2} + e^{ \frac{(\beta_{1}- \beta_{2} )}{3} \phi(\omega) } ( - d t^{2} + d x^{2} + d y^{2} + d z^{2} )
\label{60}
\end{equation}
with
\begin{equation}
\phi(\omega) = \frac{ 6 }{4 \beta_{2} - \beta_{1} } \ln \left\vert \left( \frac{4 \beta_{2} - \beta_{1} }{6} \right) \sqrt{ \frac{ 6 \left( \frac{ \beta_{2} \beta }{2} Q^{2} + \Lambda \beta_{1} \right) }{ ( \beta_{1} - 4 \beta_{2} ) \alpha } } \omega + C \right\vert
\label{61}
\end{equation}
provided that the constants satisfy
\begin{equation}
( \beta_{1} - \beta_{2} ) \left( \frac{ \beta Q^{2} \beta_{2} }{2} + \beta_{1} \Lambda \right) = \left( \frac{\beta Q^{2}}{2} + 4 \Lambda \right) \alpha .
\label{62}
\end{equation}
 $C$ is an integration constant. $H$ is given by
\begin{equation}
H = Q e^{ \frac{( \beta_{2} - \beta_{1} )}{2} \phi(\omega) } e^{1} \wedge e^{2} \wedge e^{3}
\label{63}
\end{equation}
We note that when $Q=0$ and the constants $\beta_{1}$ and $\beta_{2}$ satisfy $\beta_{1}-\beta_{2} = \frac{4 \alpha }{\beta_{1} }$,
 the solutions reduce to (\ref{58}) and (\ref{59}).

We also note that an electric dual of solutions (\ref{60}) and (\ref{61}) may be given by defining a 2-form field
\begin{equation}
 F = e^{ \beta_{2} \phi } \ast H .
\label{64}
\end{equation}
Then the solutions are identified as electrically charged
solutions.

\section{Conclusion}

\noindent We have given a class of solutions to the variational
field equations of a generalized theory of gravity in a $D$
dimensional bulk space-time derived from an  action that includes
the second-order Euler-Poincar\'{e} term and a cosmological
constant. The theory describes a heterotic type first order
effective string model  in $D$ dimensions in the Einstein frame.
The special class of plane-symmetric solutions of this model in
5-dimensions we gave refer to a 3-brane world also called a domain
wall solution in the literature \cite{lukas}.

\end{document}